\documentstyle[prc,aps,epsfig]{revtex}
\newcommand{\be}{\begin{equation}}
\newcommand{\ee}{\end{equation}}
\newcommand{\ba}{\begin{eqnarray}}
\newcommand{\ea}{\end{eqnarray}}

\begin{document}

\title{Sum rule of the correlation function}
\author{Rados\l aw Maj\footnote{Electronic address:
{\tt radcypmaj@poczta.onet.pl}}}
\address{\it Institute of Physics, \'Swi\c etokrzyska Academy \\
ul. \'Swi\c etokrzyska 15, PL - 25-406 Kielce, Poland}

\author{Stanis\l aw Mr\' owczy\' nski\footnote{Electronic address:
{\tt mrow@fuw.edu.pl}}}
\address{\it So\l tan Institute for Nuclear Studies \\
ul. Ho\.za 69, PL - 00-681 Warsaw, Poland \\
and Institute of Physics, \'Swi\c etokrzyska Academy \\
ul. \'Swi\c etokrzyska 15, PL - 25-406 Kielce, Poland}

\date{5-th December 2004}

\maketitle

\begin{abstract}

We discuss a sum rule satisfied by the correlation function of two particles 
with small relative momenta. The sum rule, which results from the completeness 
condition of the quantum states of the two particles, is derived and checked 
to see how it works in practice. The sum rule is shown to be trivially 
satisfied by free particle pairs, and then three different systems of 
interacting particles are considered. We discuss neutron and proton pairs
in the $s-$wave approximation and the case of the so-called hard spheres 
with the phase shifts taken into account up to $l=4$. Finally, the Coulomb 
system of two charged particles is analyzed. 

\end{abstract}


\section{Introduction}


The correlation functions of two identical or non-identical particles with
`small' relative momenta have been extensively studied in nuclear collisions
for bombarding energies from tens of MeV \cite{Boal:yh} to hundreds of GeV 
\cite{Heinz:1999rw}. These functions provide information about space-time 
characteristics of particle sources in the collisions. As shown by one of 
us \cite{Mrowczynski:1994rn}, the correlation function integrated over 
particle relative momentum satisfies a simple and exact relation due to 
the completeness of the particle quantum states. The relation can be used 
to get a particle phase-space density, following the method
\cite{Bertsch:qc,Brown:2000yf}, with no need to extract the Coulomb
interaction.  The sum-rule offers a constraint for the procedure
of imaging \cite{Brown:1997ku,Brown:2000aj} which inverts the correlation
function and provides the source function. The sum-rule is also helpful
to handle the correlation functions of exotic systems like $\bar p \Lambda$
\cite{Kisiel:2004it} when the interparticle correlation is poorly known.

The aim of this paper is to discuss the sum rule in detail, to prove or 
disprove its usefulness in the experimental studies. Therefore, we first 
derive the sum rule and show that it is trivially satisfied by free 
particles. Then, we analyze the correlations in the neutron-proton (n-p) 
system where there are both the two-particle scattering states and a bound 
state {\it i.e.} a deuteron. We prove that in spite of the attractive 
interaction the n-p correlation can be negative due to the deuteron 
formation. Although some qualitative features dictated by the sum rule 
are certainly seen, the calculated correlation function does not satisfy the 
sum rule. This is not surprising as not only the momenta of order $1/R$, 
where $R$ is the source size, but also larger momenta contribute to the 
sum-rule integral. Consequently, the $s-$wave approximation, which is
used compute the n-p correlation function, is insufficient.

To see how important are the higher $l$ phase shifts, we discuss a system
of the so-called hard spheres and compute the correlation function of
identical and non-identical particles, taking into account the phase shifts 
up to $l=4$. Paradoxically, the higher $l$ contributions do not improve
the situation but make it even worse.  

Finally, we discuss the correlations due to the Coulomb interaction. This 
is the case of particular interest as one usually measures the correlations 
functions of charge particles which experience the Coulomb interaction. And 
the integral, which is controlled by the sum rule, is used to determine the 
particle phase-space density \cite{Bertsch:qc,Brown:2000yf}. In the case of 
Coulomb interactions, the exact wave functions are known, and consequently, 
the exact correlation functions can be computed. Unfortunately, as we discuss 
in detail, the integral of interest appears to be divergent, and the sum rule 
does not hold for this most important case.


\section{Preliminaries}


To avoid unnecessary complications, our considerations are non-relativistic
and we start with the formula repeatedly discussed in the literature 
which expresses the correlation function $R({\bf q})$ of two particles with 
the relative momentum ${\bf q}$ as 
\be 
\label{corr-def} 
R({\bf q}) = \int d^3r \: {\cal D}_r ({\bf r}) \:
\vert \phi_{\bf q}({\bf r}) \vert^2 \;,
\ee
where $\phi_{\bf q}({\bf r})$ is the wave function of relative motion 
of the two particles and ${\cal D}_r ({\bf r})$ is the {\em effective} 
source function defined through the probability density $D_r({\bf r},t)$ 
to emit the two particles at the relative distance ${\bf r}$ and
the time difference $t$ as
\be 
\label{eff-source}
{\cal D}_r ({\bf r}) = \int dt \: D_r({\bf r} - {\bf v}t,t) \;,
\ee
with ${\bf v}$ being the particle pair velocity with respect to the
source; the relative source function $D_r({\bf r},t)$ is given by
the single-particle source function $D({\bf r},t)$, which describes
the space-time emission points of a single particle, in the following 
way
$$
D_r({\bf r},t) = \int d^3R dT \: 
D({\bf R}-{1 \over 2} {\bf r},T -{1 \over 2}t) \;
D({\bf R}+{1 \over 2} {\bf r},T +{1 \over 2}t) \;.
$$    
We note that ${\cal D}_r ({\bf r})$, $D_r({\bf r},t)$ and $D({\bf r},t)$
are all normalized as 
\be \label{norma}
\int d^3r \: {\cal D}_r ({\bf r}) = \int d^3r\, dt \: D_r({\bf r},t)
= \int d^3r\, dt \: D_r({\bf r},t) = 1 \;.
\ee 

We also observe that the spherically symmetric single-particle source 
function $D({\bf r},t)$ provides, in general, the effective 
${\cal D}_r ({\bf r})$ which is elongated along the velocity ${\bf v}$. 
To simplify calculations, we, however, assume here that the source 
function ${\cal D}_r ({\bf r})$ is spherically symmetric. Such an assumption 
makes sense when the single-particle source function $D({\bf r},t)$ 
is spherically symmetric and particles are emitted instantaneously
{\it i.e.} $D({\bf r},t) = D(|{\bf r}|) \, \delta(t-t_0)$. Then,
$$
{\cal D}_r ({\bf r}) =  \int d^3R \:
D({\bf R}-{1 \over 2} {\bf r}) \;
D({\bf R}+{1 \over 2} {\bf r}) \;.
$$ 
The single-particle source function is often chosen in the Gaussian
form
$$ 
D({\bf r}) = {1 \over (2 \pi r_0^2)^{3/2}}\: 
{\rm exp} \Big(-{{\bf r}^2 \over 2r_0^2}\Big) \;.
$$
It gives the mean radius squared of a source equal  
$\langle {\bf r}^2 \rangle = 3 r_0^2$, and it leads to the effective
relative source function as
\be 
\label{gauss}
{\cal D}_r({\bf r}) = {1 \over (4 \pi r_0^2)^{3/2}}\:
{\rm exp} \Big(-{{\bf r}^2 \over 4r_0^2}\Big) \;.
\ee


\section{Sum Rule}


Let us  consider the correlation function integrated over
the relative momentum. Since $R({\bf q}) \rightarrow 1$ when
${\bf q} \rightarrow \infty$, we rather discuss the integral
of $R({\bf q}) - 1$. Using Eq.~(\ref{corr-def}) and taking
into account the normalization condition of ${\cal D}_r ({\bf r})$
(\ref{norma}), one finds, after changing order of the ${\bf r}-$ and 
${\bf q}-$integration, the expression 
\be \label{sum1}
\int {d^3 q \over (2\pi )^3} \; \Big( R({\bf q}) - 1 \Big)
= \int d^3r \;  {\cal D}_r ({\bf r}) \;
\int {d^3 q \over (2\pi )^3} \;
\Big( \vert \phi_{\bf q}({\bf r}) \vert ^2 - 1 \Big) \;.  
\ee

It appears that the integral over ${\bf q}$ in the r.h.s. of 
Eq.~(\ref{sum1}) is determined by the quantum-mechanical completeness 
condition. Indeed, the wave functions satisfy the well-known closure
relation
\be \label{complete1}
\int {d^3 q \over (2\pi )^3} \;
\phi_{\bf q}({\bf r}) \phi^*_{\bf q}({\bf r}') +
\sum_{\alpha} \phi_{\alpha}({\bf r}) \phi^*_{\alpha}({\bf r}')
= \delta^{(3)}({\bf r} - {\bf r}') \pm \delta^{(3)}({\bf r} + {\bf r}')
\;,  
\ee where $\phi_{\alpha}$ represents a possible bound state of the two 
particles of interest. When the particles are not identical the second 
term in the r.h.s. of Eq.~(\ref{complete1}) is not present. This term 
guarantees the right symmetry for both sides of the equation for the case 
of identical particles. The upper sign is for bosons while the lower one 
for fermions. The wave function of identical bosons (fermions) 
$\phi_{\bf q}({\bf r})$ is (anti-)symmetric when 
${\bf r} \rightarrow -{\bf r}$, and the r.h.s of Eq.~(\ref{complete1}) 
is indeed (anti-)symmetric when ${\bf r} \rightarrow -{\bf r}$ or 
${\bf r}' \rightarrow -{\bf r}'$. If the particles of interest carry 
spin, the summation over spin degrees of freedom in the l.h.s. of 
Eq.~(\ref{complete1}) is implied. When the integral representation of 
$\delta^{(3)}({\bf r} - {\bf r}') $ is used, the relation (\ref{sum1}) 
can be rewritten as
$$
\int {d^3 q \over (2\pi )^3} \; \Big(
\phi_{\bf q}({\bf r}) \phi^*_{\bf q}({\bf r}') 
- e^{i{\bf q}({\bf r}-{\bf r}')} \Big) 
+ \sum_{\alpha} \phi_{\alpha}({\bf r}) \phi^*_{\alpha}({\bf r}') 
= \pm \delta^{(3)}({\bf r} + {\bf r}') \;.
$$
Now, we take the limit ${\bf r}' \rightarrow {\bf r}$ and get the relation 
\be 
\label{complete2} 
\int {d^3 q \over (2\pi )^3} \;
\Big( \vert \phi_{\bf q}({\bf r}) \vert ^2 - 1 \Big)
=  \pm \; \delta^{(3)}(2 {\bf r})
- \sum_{\alpha} \vert \phi_{\alpha}({\bf r}) \vert ^2 \;. 
\ee
When Eq.~(\ref{complete2}) is substituted into Eq.~(\ref{sum1}), 
one finds the desired sum rule
\be 
\label{sum}
\int d^3 q \; \Big( R({\bf q}) - 1 \Big)
= \pm \pi^3 \:  {\cal D}_r (0)
- \sum_{\alpha} A_{\alpha} \;. 
\ee
where $A_{\alpha}$ is the formation rate of a bound state
${\alpha}$ 
\be \label{form-rate}
A_{\alpha} =  (2\pi)^3 \int d^3r \:
{\cal D}_r({\bf r}) \vert \phi_{\alpha}({\bf r}) \vert ^2 \;.
\ee
$A_{\alpha}$ relates the cross section to produce the bound state ${\alpha}$
with the momentum ${\bf P}$ to the cross section to produce the two 
particles with the momenta ${\bf P}/2$ as
$$
{d \sigma ^{\alpha} \over d{\bf P}} = A_{\alpha} \;
{ d \;\widetilde \sigma  \over d({\bf P}/2) d({\bf P}/2) } \;.
$$
The tilde means that the short range correlations are removed from the 
two-particle cross section which is usually taken as a product of the 
single-particle cross sections.

The completeness condition is, obviously, valid for any inter-particle 
interaction. It is also valid when the pair of particles interact with 
the time-independent external field, {\it e.g.} the Coulomb field generated 
by the particle source. Thus, the sum rule (\ref{sum}) holds under very 
general conditions as long as the basic formula (\ref{corr-def}) is 
justified, in particular as long as the source function ${\cal D}_r ({\bf r})$ 
is ${\bf q}-$independent and spin independent. The validity of these 
assumptions can be only tested within a microscopic model of nucleus-nucleus 
collision which properly describes the quantum particle correlations and
bound state formation. 

There are several potential applications of the relation (\ref{sum}). 
As already mentioned, the integral of the correlation function in the 
l.h.s. of Eq.~(\ref{sum}) is used to determine a particle phase-space 
density. The method \cite{Bertsch:qc,Brown:2000yf} assumes that the 
correlation function represents non-interacting particles. Consequently,
before computing the integral, the correlation function is corrected 
in such a way that the Coulomb interaction is removed. However, the sum 
rule (\ref{sum}) shows that the integral of the correlation function 
is independent of the inter-particle interaction. Therefore, there is 
no need to extract the Coulomb interaction. We note that this procedure 
is rather model dependent.

To get information about the source function $D({\bf r},t)$, one
usually parameterizes the function, computes the correlation function
and compares it with experimental data. The method of imaging 
\cite{Brown:1997ku,Brown:2000aj} provides the source function directly, 
inverting the functional $R[{\cal D}_r]$ (\ref{corr-def}) with the 
experimental correlation function as an input. As seen, the relation
(\ref{sum}), which gives ${\cal D}_r (0)$, can be treated as a useful 
constraint of the imaging method.

The sum-rule is also helpful to better understand the correlation 
functions. In particular, the sum rule shows that the correlation
function can be negative in spite of an attractive inter-particle 
interaction. This happens when the particles form bound states 
represented by the second term in the r.h.s. of Eq.~(\ref{sum}) or 
the interaction is strongly absorptive as in the case of antiproton-lambda 
system which annihilates into mesons. Then, the scattering states of
the $\bar p \Lambda$ do not give a complete set of quantum states 
of the system and the integral from the l.h.s. of Eq.~(\ref{sum})  
has to be negative. The recently measured correlation function of 
$\bar p$ and $\Lambda$ \cite{Kisiel:2004it} is indeed negative at 
small relative momenta.  

 
\section{Free particles}


In the case of non-interacting particles, the correlation function differs 
from unity only for identical particles. Then, the wave function, which 
enters the correlation function (\ref{corr-def}), is an (anti-)symmetrized 
plane wave
$$
\phi_{\bf q}({\bf r}) = \frac{1}{\sqrt{2}}\Big(
e^{i{\bf q r}} \pm e^{-i{\bf q r}} \Big) \;,
$$
with the upper sign for bosons and lower for fermions. Then, the integration
over ${\bf q}$ in Eq.~(\ref{sum1}) can be explicitly performed without 
reference to the completeness condition (\ref{complete2}), and one finds
\be 
\label{sum-free}
\int d^3 q \; \Big( R({\bf q}) - 1 \Big)
= \pm \pi^3 \:  {\cal D}_r (0) \;. 
\ee
It this way the sum rule (\ref{sum-free}) was found in \cite{Podgoretsky:ut}, 
see also \cite{Bertsch:qc}.

Although the sum rule (\ref{sum}) assumes the integration up to the infinite 
momentum, one expects that the integral in Eq.~(\ref{sum}) saturates 
at sufficiently large $q$. To discuss the problem quantitatively, we define 
the function
\be 
\label{smax}
S(q_{\rm max}) = 4\pi \int_0^{q_{\rm max}} dq \: q^2 (R(q) - 1) \;.
\ee
As already mentioned, the source function is assumed to be spherically
symmetric, and consequently the correlation function depends only on 
$q \equiv |{\bf q}|$. Thus, the angular integration is trivially performed. 

For the Gaussian effective source (\ref{gauss}), when 
$\pi^3 {\cal D}_r (0) = (\sqrt{\pi}/2r_0)^3$, the free functions
$R(q)$ and $S(q_{\rm max})$ equal
$$
R(q) = 1 \pm e^{- 4 r_0^2q^2} \;,
$$
$$
S(q_{\rm max}) = \pm  \bigg( \frac{\sqrt{\pi}}{2r_0} \bigg)^3
{\rm E}_{2/3}\big( (2r_0 q_{\rm max})^3 \big) \;,
$$
where
$$
{\rm E}_n(x) \equiv \frac{1}{\Gamma (1 + 1/n)} 
\int_0^x dt \: e^{-t^n} \;,
$$
and $\Gamma (z)$ is the Euler gamma function. Since for large $x$, the 
function ${\rm E}_n(x)$ can be expressed as, see {\it e.g.} \cite{Janke60},
$$
{\rm E}_n(x) = 1 - \frac{1}{\Gamma (1 + 1/n)} 
\: \frac{e^{-x^n}}{n x^{n-1}} \: 
\bigg( 1 + {\cal O}\Big(\frac{1}{x^{n}} \Big) \bigg) \;,
$$
we have the approximation
$$
S(q_{\rm max}) \approx \pm  \bigg( \frac{\sqrt{\pi}}{2r_0} \bigg)^3
\Big( 1 - \frac{4r_0 q_{\rm max}}{\sqrt{\pi}} \, 
e^{- 4 r_0^2 \, q^2_{\rm max}}  \Big) \;,
$$
for $(2r_0 \, q_{\rm max})^3 \gg 1$. As seen, the sum-rule integral 
(\ref{sum-free}) is saturated for $q_{\rm max}$ not much exceeding 
$1/2r_0$, and obviously, the sum rule is satisfied.

 
\section{Neutron-Proton system}


In this section, we discuss the interacting neutron-proton pair which is 
either in the spin singlet or triplet state. The nucleons, produced in 
high-energy nuclear collisions, are usually assumed to be unpolarized, 
and one considers the spin-averaged correlation function $R$ which is 
a sum of the singlet and triplet correlation functions $R^{s,t}$ with 
the weight coefficients 1/4 and 3/4, respectively. Here, we consider, however, 
the singlet and the triplet correlation functions separately. Then, 
the sum rule (\ref{sum}) reads
\ba 
\label{sum-s}
\int d^3 q \; \Big( R^s({\bf q}) - 1 \Big)
&=& 0 \;, 
\\ [2mm] \label{sum-t}
\int d^3 q \; \Big( R^t({\bf q}) - 1 \Big)
&=& - A_d \;.
\ea

Following \cite{Lednicky82}, we calculate the correlation functions
$R^{s,t}$ assuming that the source radius is significantly larger
than the n-p interaction range. Then, the wave function of the n-p 
pair (in a scattering state) can be approximated by its asymptotic form 
\be 
\label{scatter}
\phi _{np}^{s,t}({\bf r}) = e^{i{\bf q}{\bf r}}  + f^{s,t}(q)
{e^{iq r} \over r  } \;, 
\ee
where $f^{s,t}({\bf q})$ is the scattering amplitude. It is chosen as
\be \label{ampli}
f^{s,t} (q)  = { -a^{s,t} \over 1 - {1 \over 2} d^{s,t}a^{s,t}q^2 
+ i q a^{s,t} } \;, 
\ee 
where $a^{s,t}$ ($d^{s,t}$) is the scattering length (effective range) 
of the n-p scattering; $a^s = - 23.7$ fm, $d^s = 2.7$ fm and $a^t =  5.4$ fm, 
$d^t = 1.7$ fm \cite{McCarthy68}. The amplitude (\ref{ampli}) takes into 
account only the $s-$wave scattering. This is justified as long as only 
`small' relative momenta are considered.

Substituting the wave function (\ref{scatter}) into the formula 
(\ref{corr-def}) with the source function (\ref{gauss}), we get 
\ba \label{n-p-correl}
R^{s,t} (q) =  1 +  { \rm Re} \big[ f^{s,t}(q) \big] \:
{1 \over 2 r_0^2 q} \: e^{- 4 r_0^2 q^2 } \; {\rm erfi}(2 r_0 q) 
- { \rm Im} \big[ f^{s,t}(q) \big] \:
{1 \over 2 r_0^2 q} \big( 1 - e^{- 4 r_0^2 q^2 } \big) 
+ |f^{s,t}(q)|^2 { 1 \over 2 r_0^2 } \;,
\ea
where
$$
{\rm erfi}(x) \equiv {2 \over \sqrt{ \pi}} \int_0^x dt e^{t^2} \;.
$$
Since the source described by the formula (\ref{gauss}) is spherically 
symmetric, the correlation function (\ref{n-p-correl}) does not depend 
on ${\bf q}$ but on $q$ only.

\begin{figure}
\hspace{2cm}
\epsfig{file=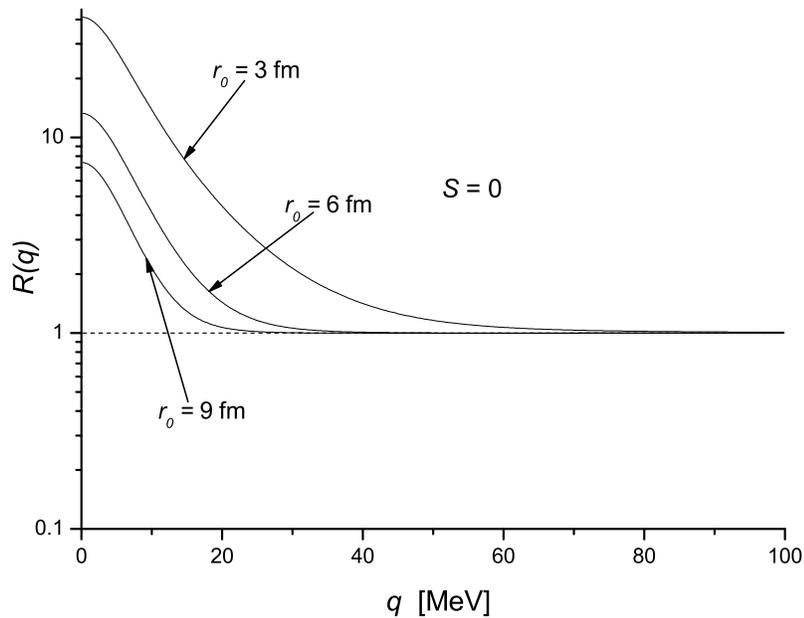,width=120mm}

\caption{\label{n-p-singlet-corr}
The singlet correlation function of neutron and proton computed
for three values of the source size parameter $r_0$.}

\end{figure}

\begin{figure}

\hspace{2cm}
\epsfig{file=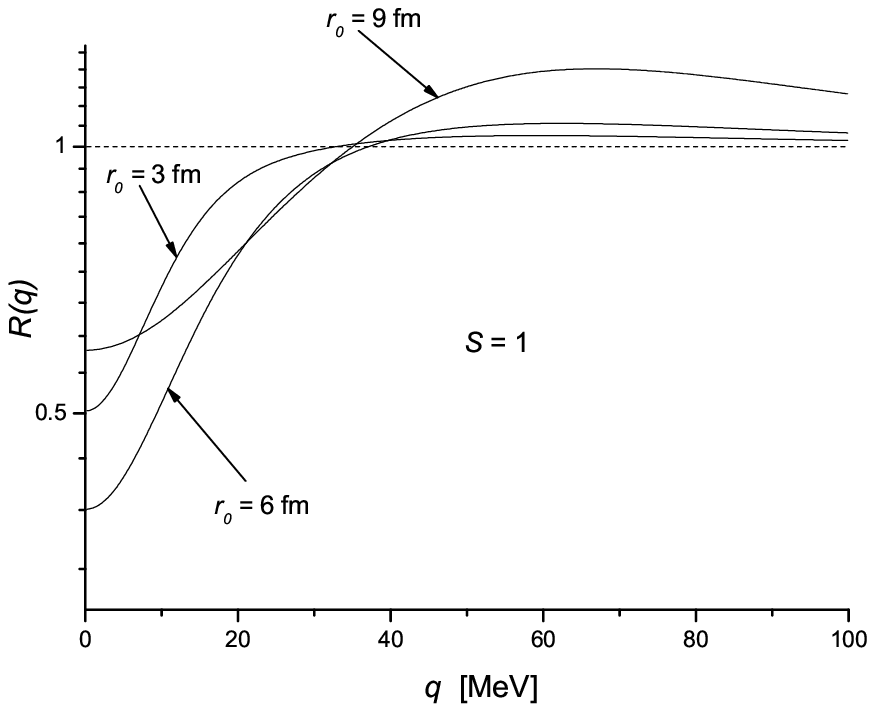,width=120mm}

\caption{\label{n-p-triplet-corr}
The triplet correlation function of neutron and proton computed
for three values of the source size parameter $r_0$.}

\end{figure}

In Figs.~\ref{n-p-singlet-corr} and \ref{n-p-triplet-corr} we show, 
respectively, the singlet and triplet correlation functions computed 
for three values of the source size parameter $r_0$. As seen, the triplet 
correlation is negative in spite of the attractive neutron-proton 
interaction. This happens, in accordance with the sum rule (\ref{sum-t}), 
because the neutron and proton, which are close to each other in the 
phase-space, tend to exist in a bound not in a scattering state.
And the n-p pairs, which form a deuteron, deplete the sample of n-p
pairs and produce a dip of the correlation function at small relative
momenta.

\begin{figure}

\hspace{2cm}
\epsfig{file=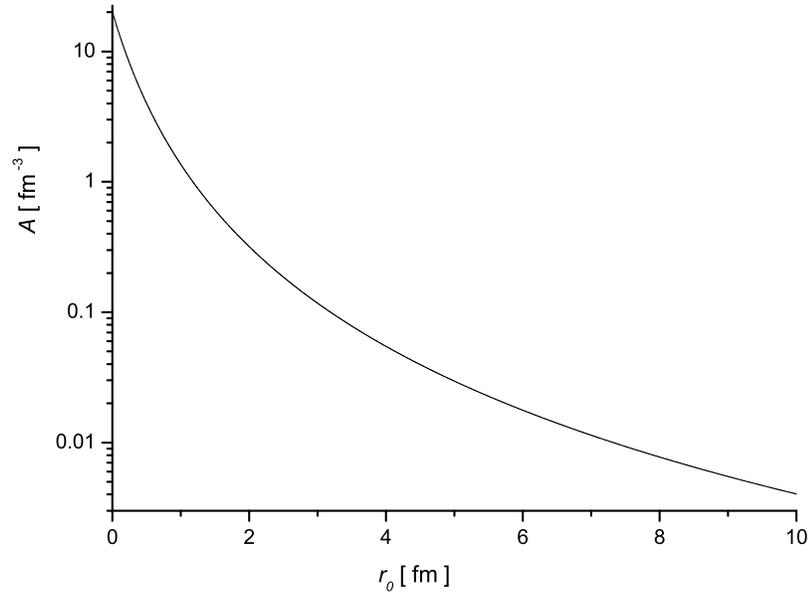,width=120mm}

\caption{\label{n-p-d-rate}
The deuteron formation rate as a function of the source size 
parameter $r_0$.}

\end{figure}

The deuteron formation rate, which enters the sum rule (\ref{sum-t}), is 
computed with the deuteron wave function in the Hulth\' en form 
\be \label{Hulthen} 
\phi _d({\bf r}) = \Big ( 
{\alpha \beta (\alpha + \beta) \over 2\pi (\alpha - \beta )^2} \Big )^{1/2} 
\;\; {e^{-\alpha r}-e^{-\beta r}  \over r } \;, 
\ee
with $\alpha = 0.23$ fm$^{-1}$ and $\beta = 1.61$ fm$^{-1}$ 
\cite{Hodgson71}. Substituting the wave function (\ref{Hulthen})
and the source function (\ref{gauss}) into Eq.~(\ref{form-rate}),
we get
\be \label{d-rate}
A_d = {2 \pi^2 \over r_0^2} \;
{\alpha \beta (\alpha + \beta ) \over  (\alpha - \beta )^2 } 
\big[ K(2\alpha r_0) - 2 K((\alpha + \beta ) r_0) + K( 2\beta r_0) \big] \;, 
\ee 
where
$$
K(x) \equiv  e^{x^2} {\rm erfc}(x) \;,
\;\;\;\;\;\;\;\;\;\;
{\rm erfc}(x) \equiv {2 \over \sqrt{ \pi}} \int_x^{\infty} dt e^{-t^2} \;.
$$
In Fig.~\ref{n-p-d-rate} we present the deuteron formation rate 
(\ref{d-rate}) as function of the source size parameter $r_0$. As seen, 
$A_d$ monotonously decreases when the source grows.

\begin{figure}

\hspace{2cm}
\epsfig{file=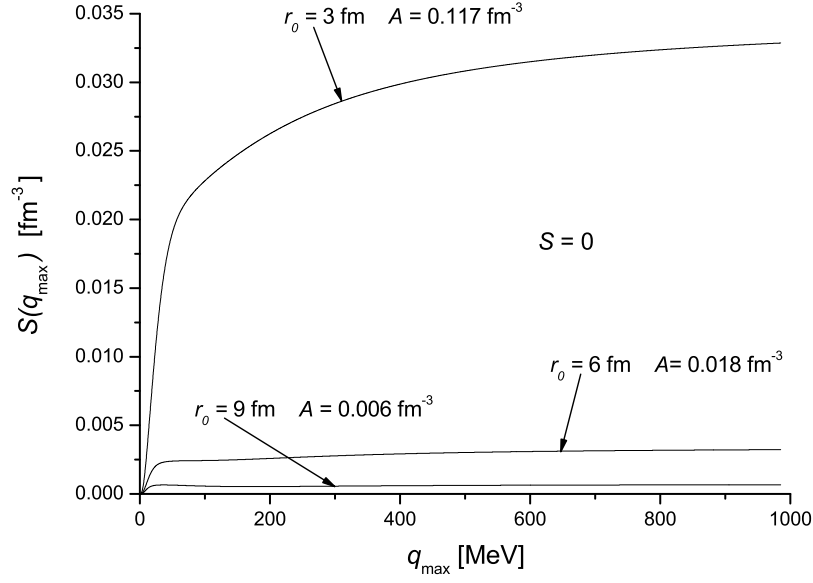,width=120mm}

\caption{\label{n-p-singlet-s}
The function $S(q_{\rm max})$ for to the singlet correlation function 
of neutron and proton for three values of the source size parameter 
$r_0$. To set the scale the corresponding values of the deuteron 
formation rate are given.}

\end{figure}

\begin{figure}

\hspace{2cm}
\epsfig{file=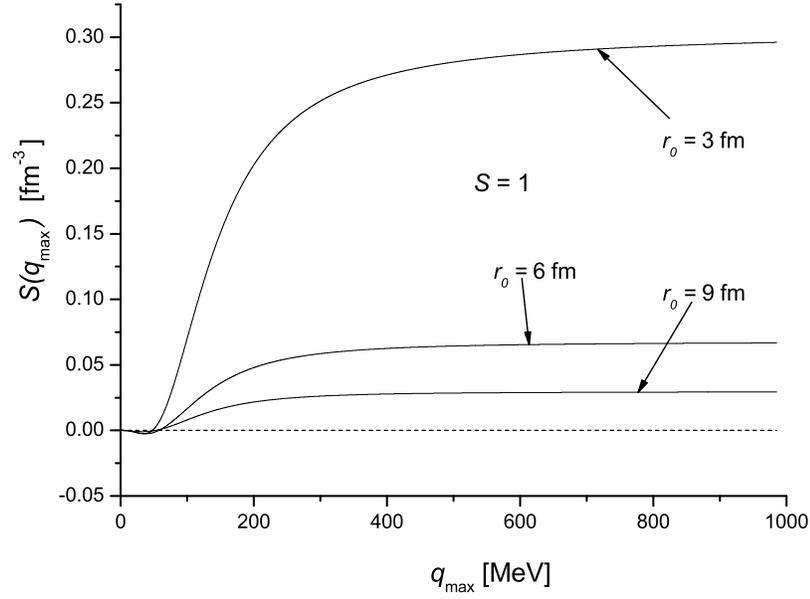,width=120mm}

\caption{\label{n-p-triplet-s}
The function $S(q_{\rm max})$ for to the triplet correlation function 
of neutron and proton for three values of the source size parameter $r_0$.}

\end{figure}

In Figs.~\ref{n-p-singlet-s} and \ref{n-p-triplet-s} we display the 
function $S(q_{\rm max})$, defined by Eq.~(\ref{smax}), found for the 
singlet and triplet correlation functions presented in 
Figs.~\ref{n-p-singlet-corr} and \ref{n-p-triplet-corr}, respectively. 
Although $S(q_{\rm max})$  saturates at large $q_{\rm max}$ for both the 
singlet and for the triplet correlation function, neither the sum rule 
(\ref{sum-s}) nor (\ref{sum-t}) is satisfied. The singlet $S(q_{\rm max})$ 
does not vanish at large $q_{\rm max}$, as it should according to
Eq.~(\ref{sum-s}), while the triplet $S(q_{\rm max})$ is not negative,
as required by Eq.~(\ref{sum-t}). However, comparing the numerical values 
of $S(q_{\rm max})$ to the corresponding deuteron formation rate, 
which sets the characteristic scale, one sees that with growing $r_0$, 
the sum rule is violated less and less dramatically. It is not surprising 
as the asymptotic form of the wave function (\ref{scatter}) and 
the $s-$wave approximation become then more accurate. The formula 
(\ref{smax}) shows that due to the factor $q^2$ even small deviations 
of the correlation function from unity at large $q$ generate sizable 
contribution to $S$. Therefore, it can be expected that higher partial 
waves have to be taken into account to satisfy the sum rule.


\section{Hard spheres}


We consider here a pair of the so-called hard spheres interacting
via the potential which is zero at distances larger than $a$ and
it is infinite otherwise; $a$ is the sphere's diameter. As well known, 
see {\it e.g.} \cite{Schiff68}, scattering of the spheres is analytically 
tractable and the $l-$th phase shift $\delta_l$ is given by the equation 
\be
{\rm tg}\delta_l = \frac{j_l(qa)}{n_l(qa)} \;,
\ee
where $j_l(z)$ and $n_l(z)$ is the spherical Bessel and Neumann function,
respectively.  The scattering amplitude equals
\be
\label{ampli-hard}
f(\Theta) =
{1 \over q}\sum_{l=0}^{\infty}(2l +1)\: P_l({\rm cos}\Theta)
\frac{{\rm ctg}\delta_l + i}{{\rm ctg}^2\delta_l+1} \;,
\ee
where $\Theta$ is the scattering angle and  $P_l(z)$ is the $l-$th
Legendre polynomial. Substituting the amplitude (\ref{ampli-hard})
into the equation analogous to Eq.~(\ref{scatter}), we get the
(asymptotic) wave function. Using this function and the Gaussian 
source (\ref{gauss}), we get the correlation function of non-identical
hard spheres as
\ba
\label{hard-non-corr}
R(q) = 1 &+& \frac{1}{\sqrt{\pi}\, r_0^3 q} \sum_l (2l+1)\:
\frac{j_l(qa)}{n_l^2(qa) + j_l^2(qa)} 
\\[2mm] \nonumber 
&\times& \Bigg[\bigg( n_l(qa) \: {\rm cos}\Big(\frac{\pi l}{2} \Big)
+  j_l(qa) \: {\rm sin}\Big(\frac{\pi l}{2} \Big)\bigg)
\int_0^\infty drr \: e^{-\frac{r^2}{4r_0^2}}\:  {\rm cos}(qr) \: j_l(qr)
\\[2mm] \nonumber
&&+ \bigg( n_l(qa) \: {\rm sin}\Big(\frac{\pi l}{2} \Big)
-  j_l(qa) \: {\rm cos}\Big(\frac{\pi l}{2} \Big)\bigg)
\int_0^\infty drr \: e^{-\frac{r^2}{4r_0^2}}\:  {\rm sin}(qr) \: j_l(qr)
\Bigg]
\\[2mm] \nonumber
&+&  \frac{1}{2 r_0^2 q^2} \sum_l (2l+1)\:
\frac{j_l^2(qa)}{n_l^2(qa) + j_l^2(qa)} \;.
\ea 
For identical hard spheres, the wave function should be (anti-)symmetrized as 
${1 \over \sqrt{2}} \big(\phi_{\bf q}({\bf r}) \pm \phi_{\bf q}(-{\bf r})\big)$,
and the correlation function equals
\ba
\label{hard-id-corr}
R(q) = 1 \pm e^{- 4 r_0^2q^2} 
&+& \frac{1}{\sqrt{\pi}\, r_0^3 q} \sum_l (2l+1)\:
\frac{j_l(qa)}{n_l^2(qa) + j_l^2(qa)} \:[1 \pm (-1)^l]
\\[2mm] \nonumber
&\times& \Bigg[\bigg( n_l(qa) \: {\rm cos}\Big(\frac{\pi l}{2} \Big)
-  j_l(qa) \: {\rm sin}\Big(\frac{\pi l}{2} \Big)\bigg)
\int_0^\infty drr \: e^{-\frac{r^2}{4r_0^2}}\:  {\rm cos}(qr) \: j_l(qr)
\\[2mm] \nonumber
&&- \bigg( n_l(qa) \: {\rm sin}\Big(\frac{\pi l}{2} \Big)
+  j_l(qa) \: {\rm cos}\Big(\frac{\pi l}{2} \Big)\bigg)
\int_0^\infty drr \: e^{-\frac{r^2}{4r_0^2}}\:  {\rm sin}(qr) \: j_l(qr)
\Bigg]
\\[2mm] \nonumber
&+&  \frac{1}{2 r_0^2 q^2} \sum_l (2l+1)\:
\frac{j_l^2(qa)}{n_l^2(qa) + j_l^2(qa)} \:[1 \pm (-1)^l] \;.
\ea
We note that there is a specific asymmetry of the signs of
the corresponding terms in Eq.~(\ref{hard-non-corr}) and in 
Eq.~(\ref{hard-id-corr}). This happens because 
$((-1)^l \pm 1)/2 = \mp (-1)^l$.

\begin{figure}

\hspace{2cm}
\epsfig{file=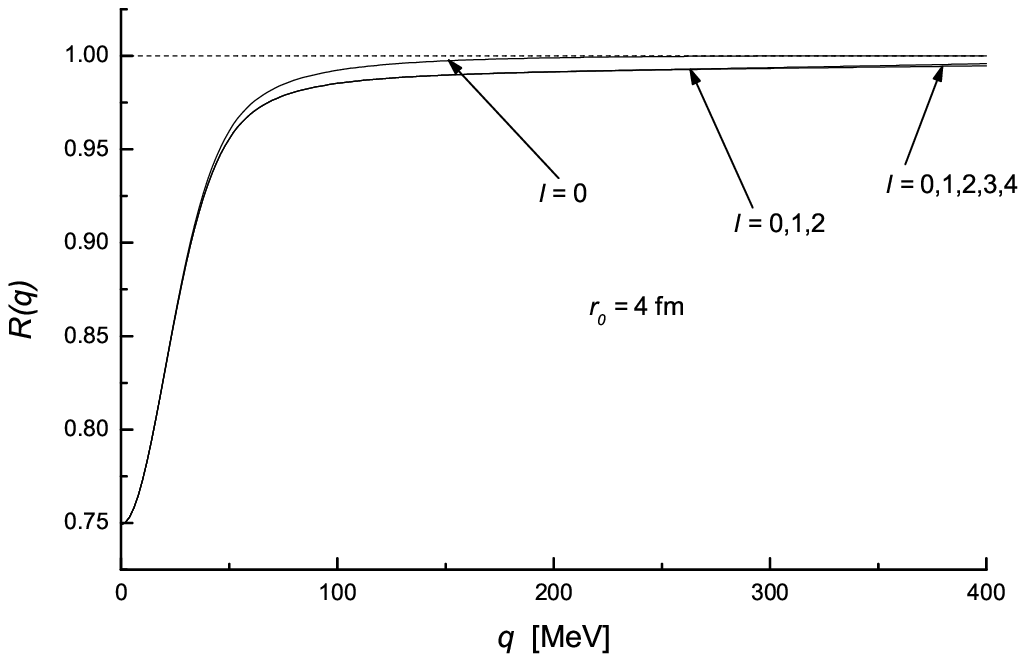,width=120mm}

\caption{\label{hard-corr-non4}
The correlation function of non-identical hard spheres for the source
size parameter $r_0 = 4$ fm.}

\end{figure}

\begin{figure}

\hspace{2cm}
\epsfig{file=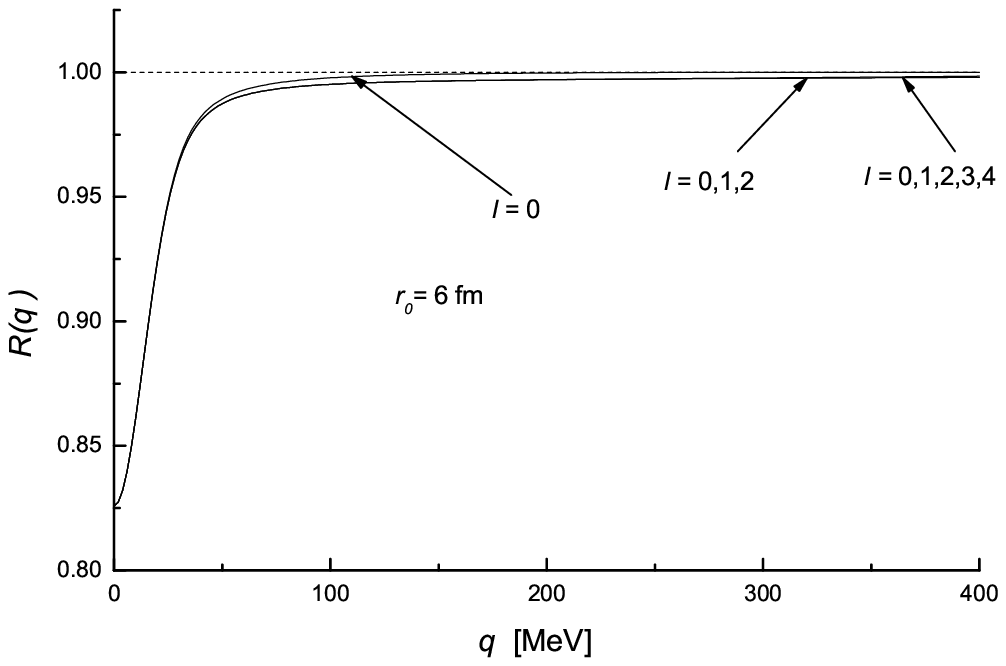,width=120mm}

\caption{\label{hard-corr-non6}
The correlation function of non-identical hard spheres for the source
size parameter $r_0 = 6$ fm.}

\end{figure}

\begin{figure}

\hspace{2cm}
\epsfig{file=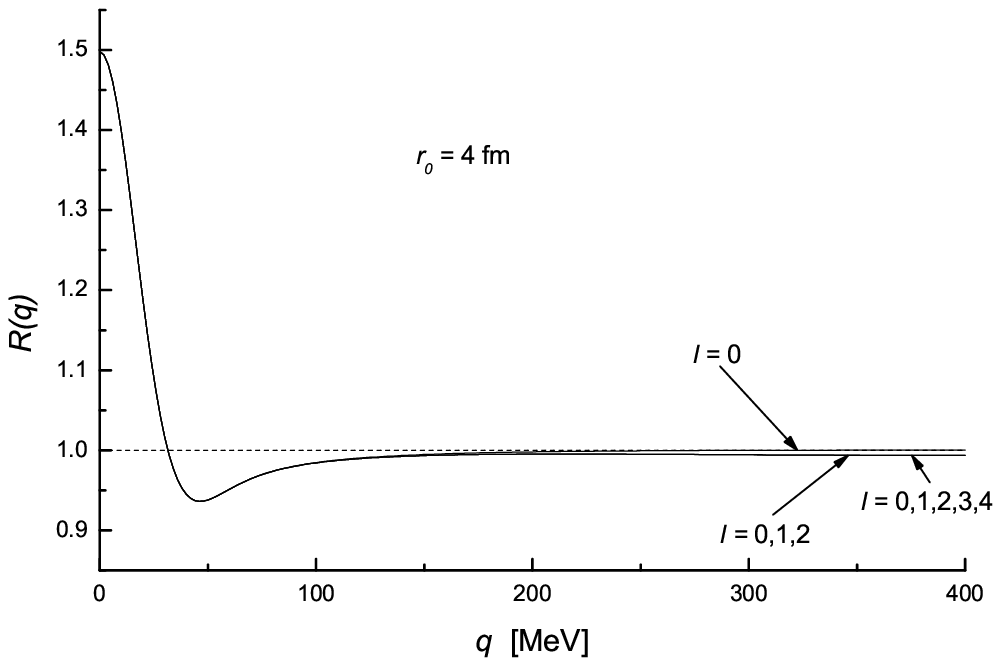,width=120mm}

\caption{\label{hard-corr-id4}
The correlation function of identical hard spheres for the source
size parameter $r_0 = 4$ fm.}

\end{figure}

\begin{figure}

\hspace{2cm}
\epsfig{file=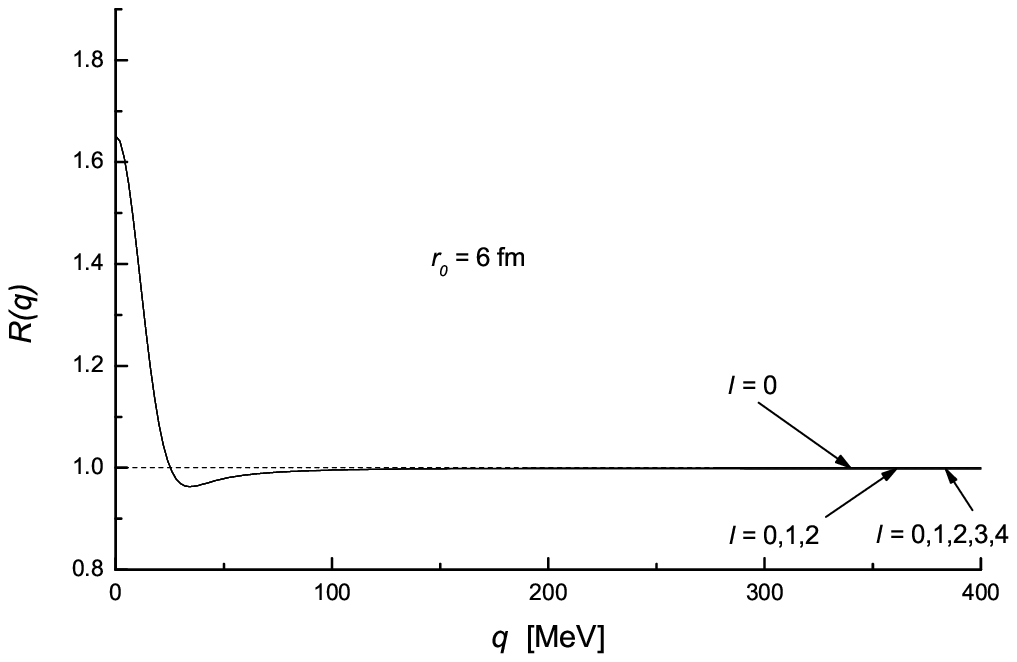,width=120mm}

\caption{\label{hard-corr-id6}
The correlation function of identical hard spheres for the source
size parameter $r_0 = 6$ fm.}

\end{figure}

\begin{figure}

\hspace{2cm}
\epsfig{file=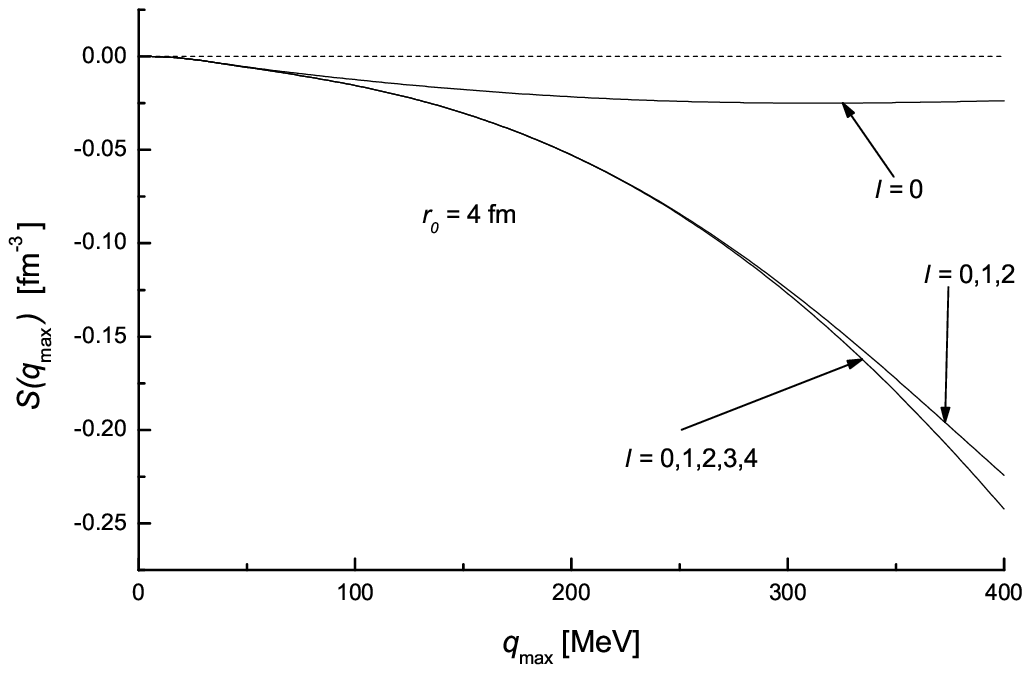,width=120mm}

\caption{\label{hard-s-non4}
The function $S(q_{\rm max})$ corresponding to the correlation function 
of non-identical hard spheres for the source
size parameter $r_0 = 4$ fm.}

\end{figure}

\begin{figure}

\hspace{2cm}
\epsfig{file=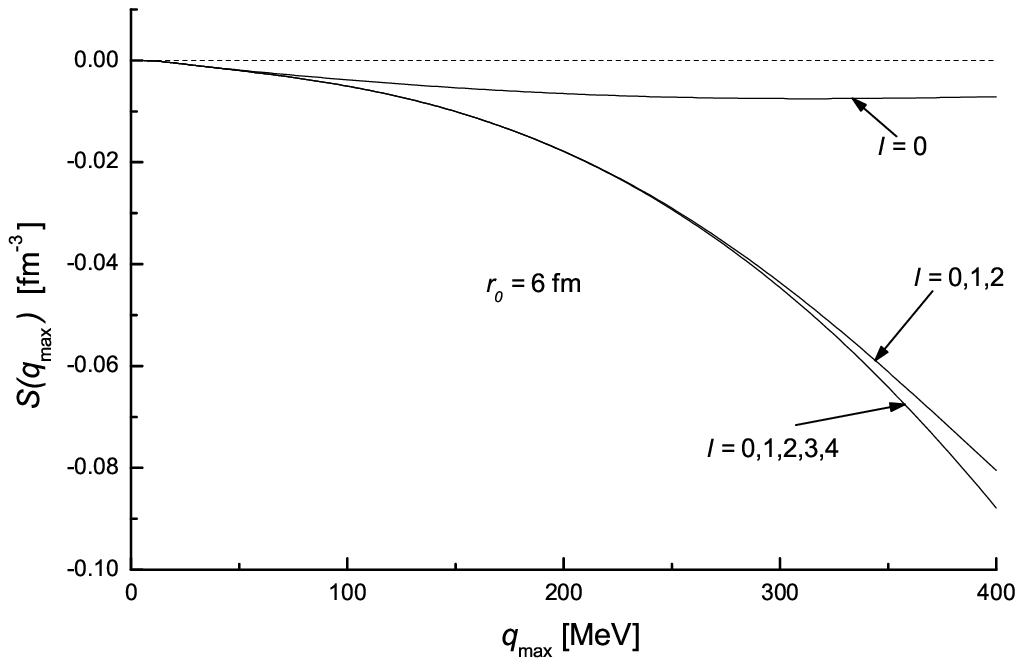,width=120mm}

\caption{\label{hard-s-non6}
The function $S(q_{\rm max})$ corresponding to the correlation function 
of non-identical hard spheres for the source
size parameter $r_0 = 6$ fm.}

\end{figure}

\begin{figure}

\hspace{2cm}
\epsfig{file=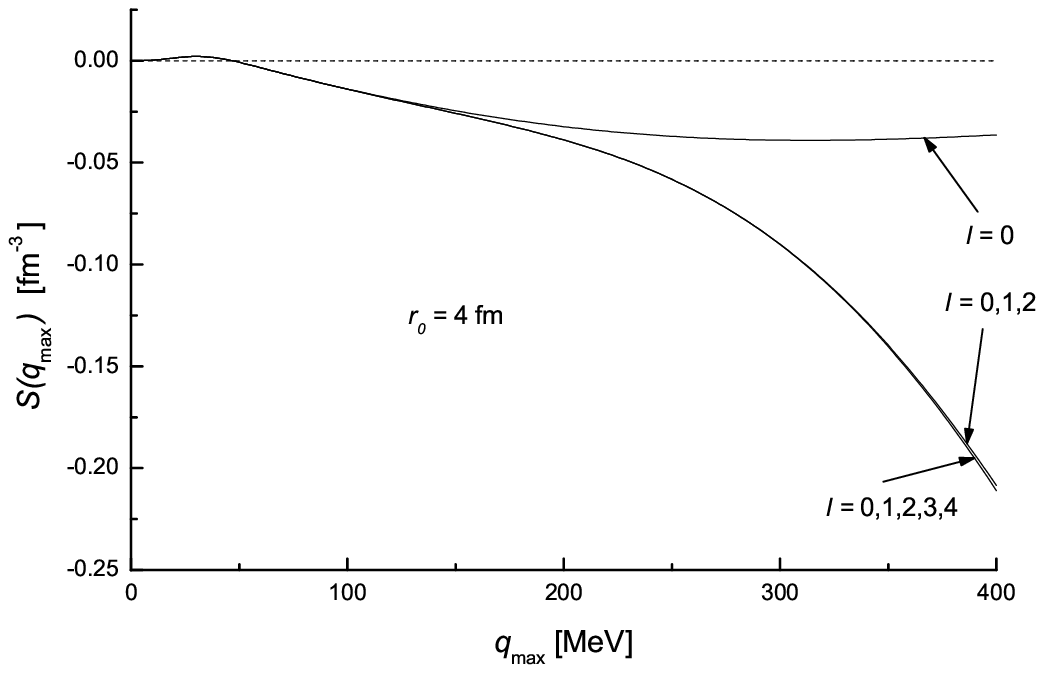,width=120mm}

\caption{\label{hard-s-id4}
The function $S(q_{\rm max})$ corresponding to the correlation function
of identical hard spheres for the source
size parameter $r_0 = 4$ fm.}

\end{figure}

\begin{figure}

\hspace{2cm}
\epsfig{file=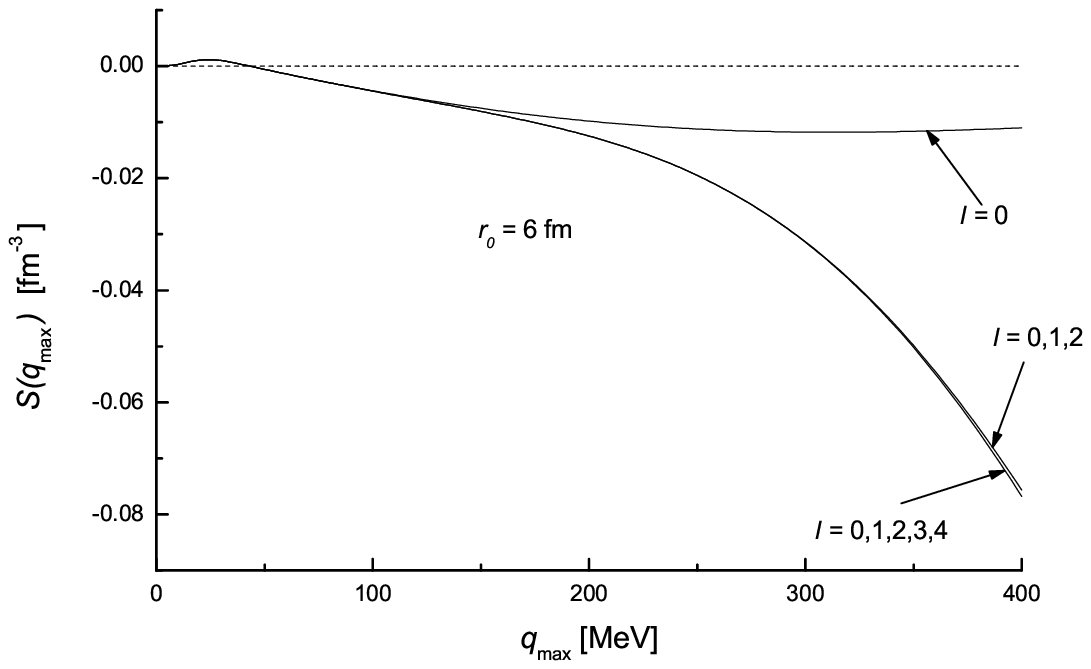,width=120mm}

\caption{\label{hard-s-id6}
The function $S(q_{\rm max})$ corresponding to the correlation function
of identical hard spheres for the source
size parameter $r_0 = 6$ fm.}

\end{figure}

In Figs.~\ref{hard-corr-non4}-\ref{hard-corr-id6} we show the correlation 
functions for non-identical spheres (\ref{hard-non-corr}) and identical 
bosonic ones (\ref{hard-id-corr}). The functions are computed numerically for 
the sphere diameter $a = 1$ fm and two values of the source size parameter 
$r_0 = 4$ fm and $r_0 = 6$ fm. We have considered rather large sources as
the wave function, which is used to compute the correlation function, is 
of the asymptotic form (\ref{scatter}). This requires $r_0 \gg a$. 
The functions $S(q_{\rm max})$ defined by Eq.~(\ref{smax}), which correspond
to the correlation functions shown in 
Figs.~\ref{hard-corr-non4}-\ref{hard-corr-id6}, are presented in 
Figs.~\ref{hard-s-non4}-\ref{hard-s-id6}. There are in 
Figs.~\ref{hard-corr-non4}-\ref{hard-s-id6} three families of curves 
representing: the $s-$wave approximation, the sum of partial waves with 
$l=0,\: 1,\: 2$, and the sum of $l=0,\: 1,\: 2,\: 3,\: 4$. As seen, going 
beyond the $s-$wave approximation minimally modifies the correlation functions 
shown in Figs.~\ref{hard-corr-non4}-\ref{hard-corr-id6}. It is rather expected
as the correlation functions are significantly different from unity only 
at small momenta where contributions of higher partial waves are strongly 
suppressed. The higher partial waves are more important for the functions
$S(q_{\rm max})$ but the integrals are still far from being saturated. 
According to the sum rule (\ref{sum}), the function $S(q_{\rm max})$ 
should tend to zero for non-identical spheres and to 
$\pi^3 {\cal D}_r (0) = (\sqrt{\pi}/2r_0)^3$ for identical ones when 
$q_{\rm max} \rightarrow \infty$. Such a behavior is not observed in our
calculations and taking into account the partial waves of higher $l$ does 
not improve the situation. We see two possible explanations of the problem. 
1) The sum rule is in principle fulfilled but one should go to much larger 
momenta $q_{\rm max}$ to saturate the integral (\ref{smax}). However, higher 
relative momenta require taking into account more and more partial waves. 
If this is indeed the case, the sum rule is formally correct but useless 
because Eq.~(\ref{corr-def}), which is a starting point of the sum rule
derivation, assumes that the relative momentum of particles $q$ 
is much smaller than the typical particle momentum.  2) The integral 
(\ref{smax}) is divergent as $q_{\rm max}$ goes to infinity. Then, the 
sum rule is simply meaningless.  We discuss the second possibility in more 
detail in the next section where the Coulomb interaction is studied.

   
\section{Coulomb interaction}


As well known, the Coulomb problem is exactly solvable within 
the non-relativistic quantum mechanics \cite{Schiff68}. The exact wave 
function of two non-identical particles interacting due to the repulsive 
Coulomb force is given as 
\be \label{coulomb-wave} \phi_{\bf q}({\bf r}) 
= e^{- {\pi \lambda \over 2 q}} \;
\Gamma (1 +i{\lambda \over q} ) \; e^{iqz/2} \; 
F(-i{\lambda \over q}, 1, iq\eta ) \;,
\ee
where $\lambda \equiv \mu e^2/8\pi$ with $\mu$ being the reduced mass 
of the two particles and $\pm e$ is the charge 
of each of them;  $F$ denotes the hypergeometric confluent function, 
and $\eta$ is the parabolic coordinate (see below). The wave function 
for the attractive interaction is obtained from (\ref{coulomb-wave}) 
by means of the substitution $\lambda \rightarrow - \lambda$. When one 
deals with identical particles, the wave function $\phi_{\bf q}({\bf r})$ 
should be replaced by its (anti-)symmetrized form. The modulus of the 
wave function (\ref{coulomb-wave}) equals
$$
|\phi_{\bf q}({\bf r})|^2 = G(q) \; 
|F(-i{\lambda \over q}, 1, iq\eta )|^2 \;,
$$
where $ G(q)$ is the so-called Gamov factor defined as
\be \label{Gamov}
G(q) = {2 \pi \lambda \over q} \, 
{1 \over {\rm exp}\big({2 \pi \lambda \over q}\big)  -1 } \;.
\ee
As seen, the modulus of the wave function of non-identical particles solely 
depends on the parabolic coordinate $\eta$. Therefore, it is natural to 
calculate the Coulomb correlation function in the parabolic coordinates: 
$\eta \equiv  r - z$, $\xi \equiv r + z$ and $\phi$ which is 
the azimuthal angle. Then, the correlation function computed for the 
Gaussian source function (\ref{gauss}) equals 
\be \label{corr-coulomb1} 
R(q) = {G(q) \over 2\sqrt{\pi} \,  r_0} \int_0^{\infty} d\eta \; 
{\rm exp}\Big(-{ \eta^2 \over 16r_0^2} \Big)  \;
|F(-i{\lambda \over q}, 1, iq\eta )|^2 \;,
\ee
where the integration over $\xi$ has been performed. It should be
noted here that, in contrast to the neutron-proton and hard sphere
cases, the Coulomb correlation function is calculated with the 
{\em exact} wave function not with the asymptotic form of it.

The modulus of the (anti-)symmetrized Coulomb wave function equals
$$
|\phi_{\bf q}({\bf r})|^2 = {1 \over 2} \; G(q) \: \Big[
|F(-i{\lambda \over q}, 1, iq\eta )|^2 
+ |F(-i{\lambda \over q}, 1, iq\xi )|^2 
\pm 2 {\rm Re}\Big(e^{iqz} \: F(-i{\lambda \over q}, 1, iq\eta ) \;
F^*(-i{\lambda \over q}, 1, iq\xi ) \Big) \Big] \;,
$$
and the correlation function analogous to (\ref{corr-coulomb1}) is
\ba \label{corr-coulomb2}
R(q) &=& {G(q) \over 2\sqrt{\pi} \, r_0} \bigg[ \int_0^{\infty} d\eta \;
{\rm exp}\Big(-{ \eta^2 \over 16r_0^2} \Big) \;
|F(-i{\lambda \over q}, 1, iq\eta )|^2  \\ [2mm] \nonumber &\pm&  {1 \over 4r_0^2} \int_0^{\infty} d\eta  \int_0^{\infty} d\xi \; 
(\eta + \xi) \; 
{\rm exp}\Big(-{ (\eta + \xi)^2 \over 16r_0^2} \Big) \;
{\rm Re}\Big(e^{iq(\xi -\eta)/2} \: 
F(-i{\lambda \over q}, 1, iq\eta ) \;
F^*(-i{\lambda \over q}, 1, iq\xi ) \Big) \bigg] \;.
\ea

\begin{figure}

\hspace{2cm}
\epsfig{file=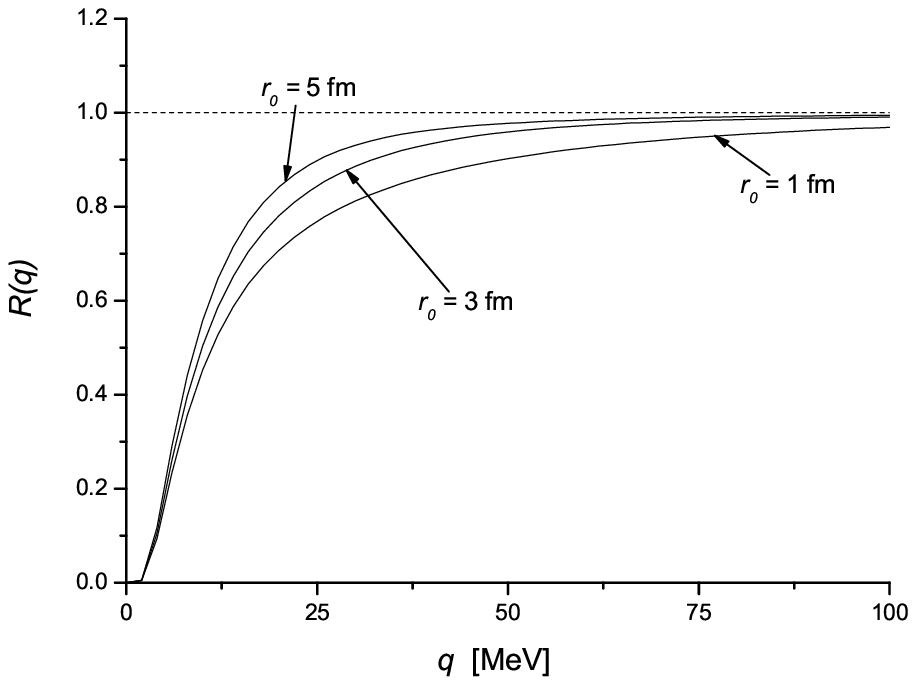,width=120mm}

\caption{\label{coulomb-non-corr}
The Coulomb correlation function of non-identical repelling 
particles for three values of the source size parameter $r_0$.}

\end{figure}

\begin{figure}

\hspace{2cm}
\epsfig{file=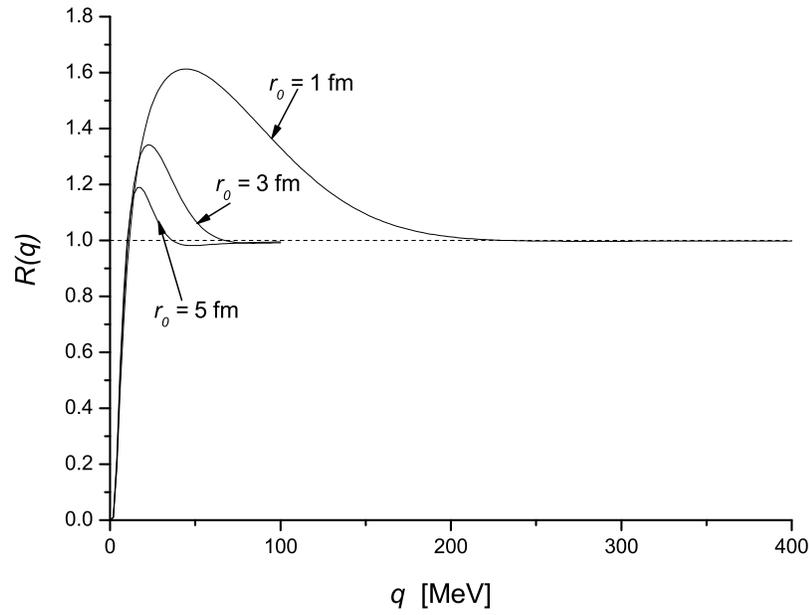,width=120mm}

\caption{\label{coulomb-id-corr}
The Coulomb correlation function of identical pions
for three values of the source size parameter $r_0$.}

\end{figure}

In Figs.~\ref{coulomb-non-corr} and \ref{coulomb-id-corr} we demonstrate 
the correlation function of non-identical particles and of identical pions, 
respectively, which are given by 
Eqs.~(\ref{corr-coulomb1},\ref{corr-coulomb2}). In the case of non-identical particles, we have taken the reduced mass, which enters the parameter 
$\lambda$, to be equal to that of identical pions. Therefore, the two 
systems differ only due to the effects of quantum statistics. In both cases 
the Coulomb interaction is repulsive. While the correlation of non-identical
particles is everywhere negative, the correlation of identical pions
is negative at small $q$ due to the Coulomb repulsion but it
is positive for larger $q$ due to the effect of quantum statistics. 

\begin{figure}

\hspace{2cm}
\epsfig{file=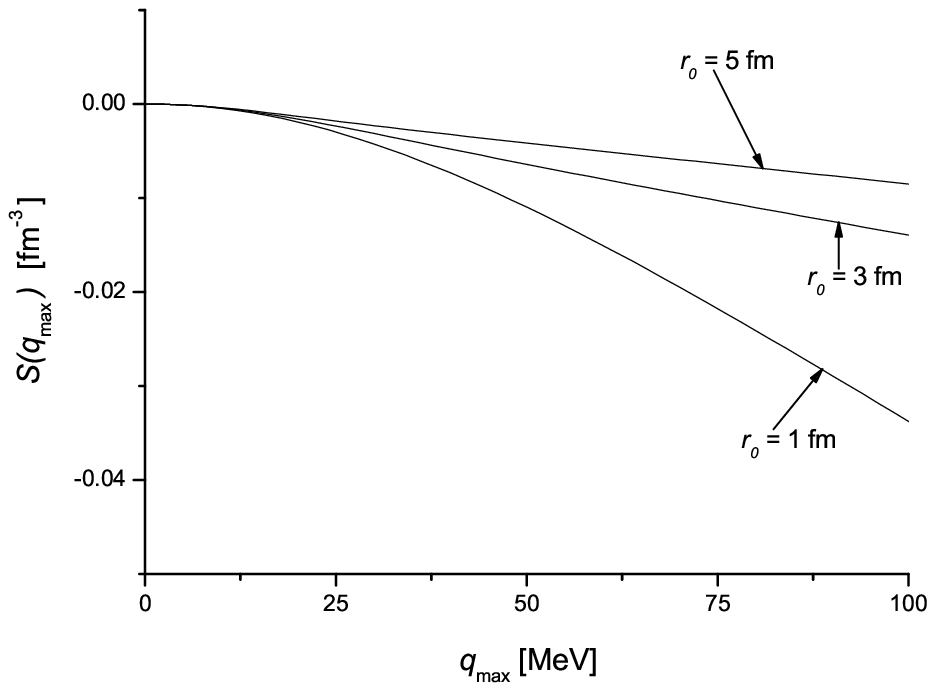,width=120mm}

\caption{\label{coulomb-non-s}
The function $S(q_{\rm max})$ corresponding to the Coulomb
correlation function of non-identical repelling particles for three 
values of the source size parameter $r_0$.}

\end{figure}

\begin{figure}

\hspace{2cm}
\epsfig{file=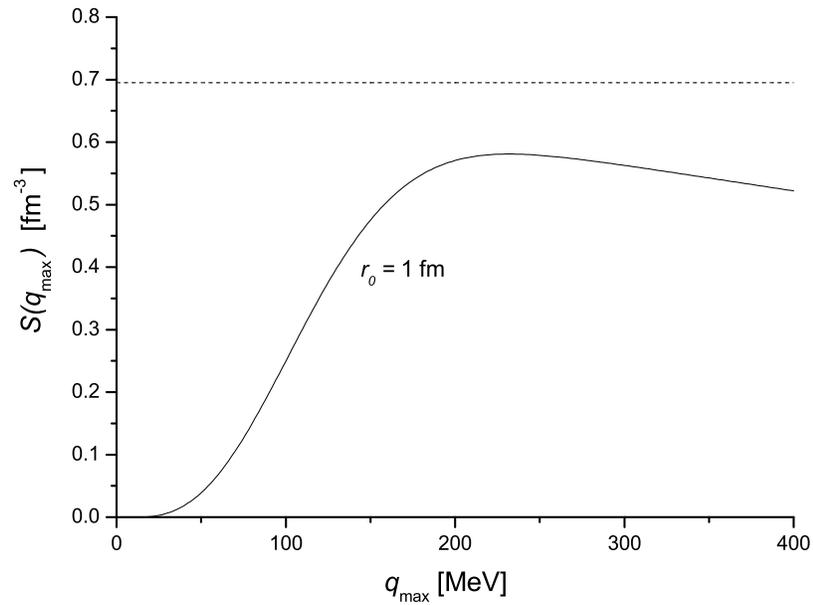,width=120mm}

\caption{\label{coulomb-id-s1}
The function $S(q_{\rm max})$ corresponding to the Coulomb
correlation function of identical pions for the source size parameter 
$r_0 = 1$ fm. The dotted line represents the value of 
$\pi^3 \, {\cal D}_r (0)$.}

\end{figure}

\begin{figure}

\hspace{2cm}
\epsfig{file=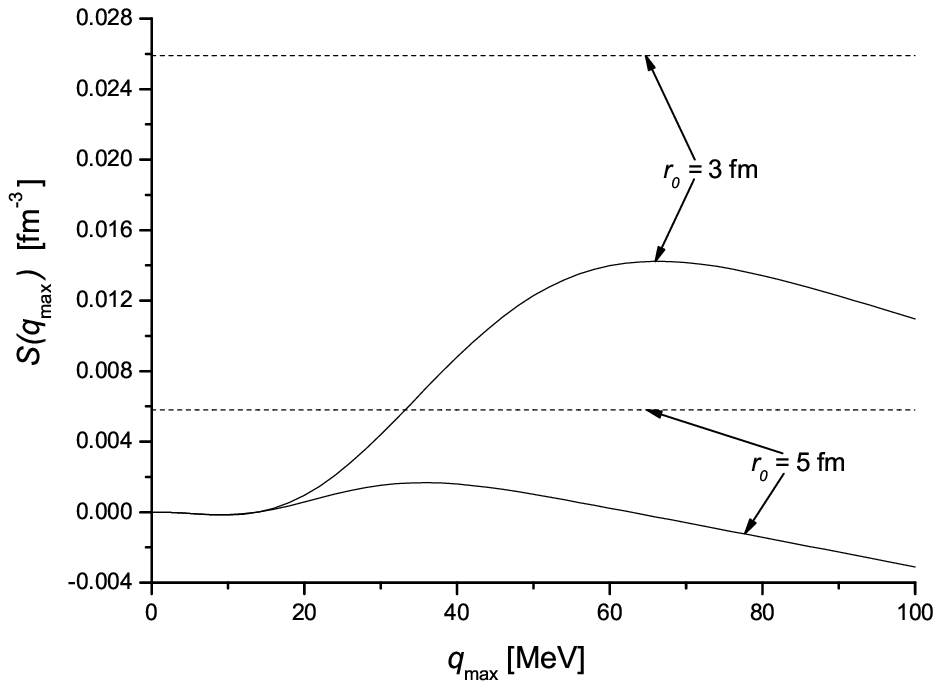,width=120mm}

\caption{\label{coulomb-id-s2}
The function $S(q_{\rm max})$ corresponding to the Coulomb
correlation function of identical pions for the source size parameter 
$r_0$ equal 3 fm and 5 fm. The dotted lines represents the respective 
values of $ \pi^3 \, {\cal D}_r (0)$.}

\end{figure}

Figs.~\ref{coulomb-non-s}, \ref{coulomb-id-s1}, and \ref{coulomb-id-s2} 
show the functions $S(q_{\rm max})$ defined by Eq.~(\ref{smax}) which 
are computed for the correlation functions presented in 
Figs.~\ref{coulomb-non-corr} and \ref{coulomb-id-corr}, respectively. 
According to the sum rule (\ref{sum}), $S(q_{\rm max})$ 
should vanish for sufficiently large $q_{\rm max}$ in the case of 
non-identical particles which repel each other, and $S(q_{\rm max})$ 
should tend to $\pi^3 \:{\cal D}_r (0)$ for identical pions.  As seen in 
Figs.~\ref{coulomb-non-s}, \ref{coulomb-id-s1}, and \ref{coulomb-id-s2}, 
the functions $S(q_{\rm max})$ do not seem to saturate at large $q_{\rm max}$ 
and the sum rule is badly violated.

What is wrong here? The derivation of the sum rule (\ref{sum})
implicitly assumes that the integral in the l.h.s. exists
{\it i.e.} it is convergent. Otherwise interchanging of the 
integrations over ${\bf q}$ and over ${\bf r}$, which leads 
to Eq.~(\ref{sum1}), is mathematically illegal. Unfortunately,
the Coulomb correlation functions appear to decay too slowly with 
$q$, and consequently the integral in the left hand side
of Eq.~(\ref{sum}) diverges.

To clarify how the integral diverges, one has to find the asymptotics
of the correlation function at large $q$. This is a difficult 
problem as one has to determine a behavior of the wave function 
(\ref{coulomb-wave}) at large $q$ for any ${\bf r}$, and then to perform 
the integration over ${\bf r}$. Up to our knowledge, the problem of
large $q$ asymptotics of the Coulomb correlation function has not
been satisfactory solved although it has been discussed in several papers  
\cite{Gyulassy:xb,Kim92,Baym:1996wk,Sinyukov:1998fc,Pratt:2003ar}.
We have not found a complete solution of the problem but our rather 
tedious analysis, which uses the analytic approximate expressions of 
the hypergeometric confluent function at small and large distances,
suggests, in agreement with 
\cite{Gyulassy:xb,Kim92,Baym:1996wk,Sinyukov:1998fc,Pratt:2003ar}, 
that $(R(q) -1) \sim 1/q^2$ when $q \rightarrow \infty$. Then, 
the integral in the left hand side of Eq.~(\ref{sum}) linearly diverges,
and the sum rule does not hold. We note that the Gamov factor (\ref{Gamov}), 
which represents a zero size source and decays as $1/q$ at large $q$, leads 
to the quadratic divergence of the integral (\ref{sum}). We also note here 
that the asymptotics $1/q^2$ of the correlation function does not have much 
to do with the well known classical limit of the correlation function 
\cite{Gyulassy:xb,Kim92,Baym:1996wk,Sinyukov:1998fc,Pratt:2003ar}. 
Since the large $q$ limit of the correlation function corresponds to the 
small separation of the charged particles, which at sufficiently large 
$q$ is smaller than the de Broglie wave length, the classical approximation 
breaks down.

   
\section{Final remarks}


The sum rule (\ref{sum}) provides a rigorous constraint on the
correlation function if the momentum integral in Eq.~(\ref{sum}) exists. 
The rule is trivially satisfied by the correlation function of
non-interaction particles. The model calculations of the n-p correlation 
function in the $s-$wave approximation fail to fulfill the sum rule.
One suspects that the approximation, which is sufficient to properly 
describe a general shape of the correlation function, distorts its tail. 
Since even small deviations of the correlation function from unity
at large $q$ generate a sizable contribution to the sum-rule integral,
it seems reasonable to expect that the higher partial waves have to be
included to comply with the sum rule. Paradoxically, when the higher 
partial waves are taken into account for the interacting hard spheres 
the situation is not improved but it gets worse. This suggests that
either one should go to really large momenta to saturate the sum-rule
integral or the integral in Eq.~(\ref{sum}) is divergent. In the case 
of Coulomb repelling interaction, we certainly deal with the second 
option. Due to the strong electrostatic repulsion at small distances, 
the correlation function decays too slowly at large momenta, and 
consequently the sum-rule integral does not exist. 

Being rather useless, the sum rule explains some qualitative features 
of the correlation function. In particular, it shows that, in spite of the 
attractive interaction, the correlation can be negative as it happens in 
the triplet state of neutron-proton pair and in the $\bar p \Lambda$
system.

\acknowledgements

We are very grateful to Wojciech Florkowski for a generous help in 
performing numerical calculations, and to David Brown, Cristina Manuel, 
Richard Lednick\'y, and Scott Pratt for fruitful discussions and 
correspondence.



\begin{thebibliography}{99}
\bibitem{Boal:yh}
D.~H.~Boal, C.~K.~Gelbke, and B.~K.~Jennings,
Rev.\ Mod.\ Phys.\  {\bf 62}, 553 (1990).

\bibitem{Heinz:1999rw}
U.~W.~Heinz and B.~V.~Jacak,
Ann.\ Rev.\ Nucl.\ Part.\ Sci.\  {\bf 49}, 529 (1999).

\bibitem{Mrowczynski:1994rn}
St.~Mr\'owczy\'nski,
Phys.\ Lett.\ B {\bf 345}, 393 (1995).

\bibitem{Bertsch:qc}
G.~F.~Bertsch,
Phys.\ Rev.\ Lett.\  {\bf 72}, 2349 (1994)
[Erratum-ibid.\  {\bf 77}, 789 (1996)].

\bibitem{Brown:2000yf}
D.~A.~Brown, S.~Y.~Panitkin, and G.~Bertsch,
Phys.\ Rev.\ C {\bf 62}, 014904 (2000)
[arXiv:nucl-th/0002039].

\bibitem{Brown:1997ku}
D.~A.~Brown and P.~Danielewicz,
Phys.\ Lett.\ B {\bf 398}, 252 (1997)
[arXiv:nucl-th/9701010].

\bibitem{Brown:2000aj}
D.~A.~Brown and P.~Danielewicz,
Phys.\ Rev.\ C {\bf 64}, 014902 (2001)
[arXiv:nucl-th/0010108].

\bibitem{Kisiel:2004it}
A.~Kisiel  [STAR Collaboration],
J.\ Phys.\ G {\bf 30}, S1059 (2004)
[arXiv:nucl-ex/0403042].

\bibitem{Podgoretsky:ut}
M.~I.~Podgoretsky,
Sov.\ J.\ Nucl.\ Phys.\  {\bf 54}, 891 (1991)
[Yad.\ Fiz.\  {\bf 54}, 1461 (1991)].

\bibitem{Janke60}
E.~Janke, F.~Emde, and F.~L\" osch, 
{\it Tafeln H\" oherer Funktionen} 
(Teubner Verlagsgesellschaft, Stuttgart, 1960).

\bibitem{Lednicky82}
R. Lednicky and V.L. Lyuboshitz, Sov. J. Nucl. Phys. {\bf 35}, 770 (1982)
[Yad. Fiz. {\bf 35}, 1316 (1982)]. 

\bibitem{McCarthy68}
I.E. McCarthy, {\it Introduction to Nuclear Theory} 
(John Wiley \& Sons, New York, 1968).

\bibitem{Hodgson71}
P.E. Hodgson, {\it Nuclear Reactions and Nuclear Structure} 
(Clarendon Press, Oxford, 1971).

\bibitem{Schiff68} L.I. Schiff, {\it Quantum Mechanics}
(McGraw-Hill, New York, 1968).

\bibitem{Gyulassy:xb}
M.~Gyulassy and S.~K.~Kauffmann,
Nucl.\ Phys.\ A {\bf 362}, 503 (1981).

\bibitem{Kim92}
Y.D.~Kim {\it et al}., Phys.\ Rev.\ C {\bf 45}, 387 (1992).
 
\bibitem{Baym:1996wk}
G.~Baym and P.~Braun-Munzinger,
Nucl.\ Phys.\ A {\bf 610}, 286C (1996).

\bibitem{Sinyukov:1998fc}
Y.~Sinyukov, R.~Lednicky, S.~V.~Akkelin, J.~Pluta, and B.~Erazmus,
Phys.\ Lett.\ B {\bf 432}, 248 (1998).

\bibitem{Pratt:2003ar}
S.~Pratt and S.~Petriconi,
Phys.\ Rev.\ C {\bf 68}, 054901 (2003).


\end{thebibliography}
\end{document}